\documentclass[english,eqsecnum,prd,aps,nofootinbib,superscriptaddress,longbibliography,tightenlines,12pt]{revtex4-2}
\usepackage[T1]{fontenc}
\usepackage[latin9]{inputenc}
\usepackage[dvipsnames]{xcolor}
\usepackage{babel}
\usepackage{mathtools}
\usepackage{amsmath}
\usepackage{amssymb,accents}
\usepackage{mathrsfs}
\usepackage{cancel}
\usepackage{comment}
\usepackage[unicode=true,pdfusetitle,
 bookmarks=true,bookmarksnumbered=false,bookmarksopen=false,
 breaklinks=true,allcolors=blue,backref=false,colorlinks=true]
 {hyperref}
\usepackage{tikz-cd}
\usepackage{tikz}
\usetikzlibrary{shapes,arrows}

\begin{document}

\title{Propagation of quantum gravity-modified gravitational waves on a classical FLRW spacetime}

\author{Angel Garcia-Chung}
\email{alechung@xanum.uam.mx} 
\affiliation{Departamento de F\'isica, Universidad Aut\'onoma Metropolitana - Iztapalapa, \\ San Rafael Atlixco 186, Ciudad de M\'exico 09340, M\'exico}
\affiliation{Universidad Panamericana, \\ Tecoyotitla 366. Col. Ex Hacienda Guadalupe Chimalistac, C.P. 01050 Ciudad de M\'exico, M\'exico}

\author{James B. Mertens}
\email{jmertens@wustl.edu}
\affiliation{Department of Physics and McDonnell Center for the Space Sciences,
Washington University, St. Louis, MO 63130, USA}
\affiliation{Department of Physics and Astronomy, York University~\\
 4700 Keele Street, Toronto, Ontario M3J 1P3 Canada}
\affiliation{Perimeter Institute for Theoretical Physics, Waterloo, Ontario N2L 2Y5, Canada}

\author{Saeed Rastgoo}
\email{srastgoo@yorku.ca}
\affiliation{Department of Physics and Astronomy, York University~\\
 4700 Keele Street, Toronto, Ontario M3J 1P3 Canada}

\author{Yaser Tavakoli}
\email{yaser.tavakoli@guilan.ac.ir}
\affiliation{Department of Physics,
University of Guilan, Namjoo Blv.,
41335-1914 Rasht, Iran}
\affiliation{School of Astronomy, Institute for Research in Fundamental Sciences (IPM),  P. O. Box 19395-5531, Tehran, Iran}

\author{Paulo Vargas Moniz}
\email{pmoniz@ubi.pt}
\affiliation{Departamento de Fisica, Centro de Matematica e Aplica\c c\~oes:  CMA-UBI, Universidade da Beira Interior, 6200 Covilh\~a,
Portugal}

\date{\today}
\begin{abstract}
The linearized Einstein field equations provide a low-energy wave equation
for the propagation of gravitational fields which may originate from
a high energy source. Motivated by loop quantum gravity, we
propose the polymer quantization scheme to derive the effective propagation of such waves on a classical 
Friedmann-Lemaitre-Robertson-Walker (FLRW) spacetime. To overcome the challenge of polymer quantizing a time-dependent Hamiltonian, we rewrite such a Hamiltonian in a time-independent manner in the extended phase space, polymerize it, and then transform it back to the usual phase space. In this way we obtain a time-dependent polymer Hamiltonian for the gravitational waves. We then derive the effective equations of motion and show that (i) the form of the waves is modified, (ii) the speed of the waves depends on their frequencies, and (iii) quantum effects become more apparent as waves traverse longer distances.
\end{abstract}
\maketitle

\section{Introduction\label{sec:Intro}}
Recent observations of gravitational waves (GWs) and the rapid increase in the sensitivity of GWs observatories has opened up a great opportunity in connecting theory and phenomenology with experiment in many areas of physics and astronomy. In particular, precision cosmology, black hole physics and quantum gravity can benefit hugely from this development. Such observations also have the potential to guide us towards the correct theory of quantum gravity by revealing the information about the deep structure of spacetime encoded in such waves. Although these effects might be extremely small, the distances these waves travel can act as amplifiers of such quantum effects, making them observable in our current or near-future experiments.

There have been numerous studies connecting theories of quantum gravity with potential observations regarding the structure of quantum spacetime. In particular, in Loop Quantum Gravity (LQG) \cite{Thiemann:2007pyv}, there have been studies to understand the consequence of nonpertubative quantization in propagation of Gamma Ray Bursts (GRBs), other matter fields, and GWs on cosmological or black holes spacetimes (for some examples see, Refs. \cite{Ashtekar:1991mz,Varadarajan:2002ht,Freidel:2003pu,Bojowald:2007cd,Ashtekar:2009mb,Hossain:2009vd,Gambini:2009ie,Mielczarek:2010bh,Date:2011bg,Gambini:2011mw,Gambini:2011nx,Sa:2011rm,Hinterleitner:2011rb,Hinterleitner:2012zz,Neville:2013wba,Neville:2013xba,Hoehn:2014qxa,Arzano:2016twc,Tavakoli:2015fvz,Dapor:2012jg, ElizagaNavascues:2016vqw,Bonder:2017ckx,Lewandowski:2017cvz,Hinterleitner:2017ard,Dapor:2020jvc,Tavakoli:2014mra,Calcagni:2020ume,Calcagni:2019kzo,Calcagni:2019ngc} and references within).
    
In this work we consider GWs as effective perturbations propagating on a classical FLRW cosmological spacetime. The effective form of such waves is derived by applying the techniques of polymer quantization \cite{Ashtekar:2002sn,Corichi:2007tf,Morales-Tecotl:2016ijb,Tecotl:2015cya,Flores-Gonzalez:2013zuk} to the classical perturbations. Such a quantization is a representation of the classical algebra on a Hilbert space that is unitarily inequivalent to the usual Schr\"{o}dinger representation. In it, operators are regularized and written in a certain exponential form. In such theories, the infinitesimal generators corresponding to some of the operators do not exist on the Hilbert space. As a consequence, the conjugate variables to those operators only admit finite transformations. Thus, the dynamics of the theory leads to the  discretization of the spectrum of the conjugate operators (for more details and some examples of polymer quantization applied to particles and path integral formulation of black holes, see Refs. \cite{garcia2014polymer, garcia2016polymer, Morales-Tecotl:2016ijb,Tecotl:2015cya,Morales-Tecotl:2018ugi}).  

Since the Hamiltonian of our model is time-dependent, we apply a certain method to overcome the challenge of polymerizing such time-dependent systems. We first write the Hamiltonian of the system in a time-independent form in the extended phase space, polymerize such a time-independent Hamiltonian, and transform it back to the usual phase space, yielding a polymerized time-dependent Hamiltonian. In fact we derive two effective Hamiltonians, each corresponding to one of the polarizations of the polymer quantization. Using these modified Hamiltonians, we study the effective equations of motion of polymerized GWs and show that i) the form of the waves is modified, ii) the speed of the waves depends on their frequencies, and iii) the quantum effects are amplified by the distance/time the waves travel.

Since the Hamiltonian of our model is time-dependent, we apply a certain method (see Ref. \cite{garcia2017dirac}) to overcome the challenge of polymerizing such a time-dependent system. We first write the system in the extended phase space formalism which turns it into a first class deparametrized system. Then, a canonical transformation in the extended phase space is applied in such a way that the constraint, or more specifically, its Hamiltonian term, in the new coordinates is time-independent. Once the constraint is solved we apply some of techniques developed \cite{austrich2017instanton, Tecotl:2015cya, Morales-Tecotl:2016dma} to obtain the polymer-quantum effective corrections to the time-independent Hamiltonian. This yields an effective (semiclassical) polymer description of the system. Consequently, we are in a position to write the system again in the extended phase space formalism and apply the inverse of the former canonical transformation. This leads us to an effective polymer-quantized time-dependent Hamiltonian representing the dynamics of the effective Fourier modes of the GWs. Moreover, we derive two effective Hamiltonians, corresponding to a discrete coordinate and a discrete momentum, respectively. Using these polymer effective Hamiltonians, we study their equations of motion which now describe the propagation of the effective GWs. We show that i) the form of the waves is modified ii) the speed of the waves depends on their frequencies, and iii) the quantum effects become more apparent as the the waves travel a longer distance/time.

This paper is organized as follows: in Sec. \ref{sec:GRW-H}, we derive the classical Hamiltonian of perturbations on an FLRW classical background. In Sec. \ref{sec:polymerization}, this time-dependent Hamiltonian is turned into a polymer effective time-dependent Hamiltonian by applying a certain method that is inspired by an approach used to deal with time-dependent Harmonic oscillators. We derive two Hamiltonians, each corresponding to one of the polarizations of the polymer quantization. In Sec. \ref{sec:phenomenology}, we derive the equations of motions corresponding to each effective polymer Hamiltonian and solve them both perturbatively and numerically in order to explore deviations from the classical behavior. Finally, in Sec. \ref{sec:conclusion} we present our concluding remarks and comment about future outlook and projects.

\section{Hamiltonian formalism for GWs\label{sec:GRW-H}}

GWs are the result of the weak-field approximation to the Einstein field
equations. On a curved spacetime, we fix the (unperturbed) background  as a 4-manifold $M = \mathbb{T}^3\times\mathbb{R}$,  with a spatial 3-torus topology\footnote{To avoid a discussion of boundary conditions on fields (generated by perturbations), we will assume that the spatial 3-manifold is $\mathbb{T}^3$.}, equipped with coordinates $x^j \in (0, \ell)$ and a temporal coordinate $x^0 \in \mathbb{R}$. We then consider a small (metric) perturbation to this background and  study the GWs generated by this perturbation.

Hence, given the (unperturbed) Einstein-Hilbert gravitational action 
\begin{equation}
S_{\rm grav}\ =\ \frac{1}{2\kappa^2}\int d^4x \sqrt{-g}\, \mathcal{R} \, ,
\label{Eq:EH-Action}
\end{equation}
the starting point of writing the Hamiltonian of the
GWs, is the general perturbed metric
\begin{equation}
g_{\mu\nu} =\ \mathring{g}_{\mu\nu} +\, h_{\mu\nu}\, ,
\label{Eq:metric-pert}
\end{equation}
where $\mathring{g}_{\mu\nu}$ is the unperturbed background metric, while $h_{\mu\nu}$ denotes a small perturbation concerning $\mathring{g}_{\mu\nu}$. Moreover,
\begin{equation}
h^{\mu\nu}\, =\, \mathring{g}^{\mu\sigma}\mathring{g}^{\nu\tau}h_{\sigma\tau}.
\end{equation}

In order to reduce the number of terms in the linearized Einstein field equations, it is convenient to express the Einstein tensor 
in terms of the {\em trace-reversed} metric perturbation
\begin{equation}
\bar{h}_{\mu\nu}\, :=\, h_{\mu\nu} - \frac{1}{2}\mathring{g}_{\mu\nu}h\, ,
\end{equation}
where 
$h=h^{~\mu}_{\mu}=\eta^{\mu\nu}h_{\mu\nu}$, with $\eta^{\mu\nu}$ being the Minkowski spacetime metric.
Thereby,  the linearized Einstein field equation in terms of  $\bar{h}_{\mu\nu}$ can be expressed as a wave equation, in the {\em Lorentz gauge}
\begin{equation}
\mathring{\nabla}_{\mu}\bar{h}^{\mu\nu}=0.
\label{LorentzGauge}
\end{equation}
Indeed, in this gauge the metric perturbation looks like a {\em transverse} wave. By imposing an additional \emph{(synchronous) transverse-traceless} gauge, where 
\begin{equation}
\bar{h}=0,\quad \quad  \bar{h}_{0\mu}=0, \quad \quad {\rm and} \quad \quad \mathring{\nabla}_{i}\bar{h}^{ij}=0,
\end{equation}
we consider only spatial, transverse, and traceless perturbations. In the latter case,  the metric perturbations $h_{ij}$ correspond at present time to   GWs propagating  on the unperturbed spacetime background. 
A wave traveling  along, say, the $x^{3}$ direction, can be separated into two polarization scalar modes $h_{+}(x)$ and $h_{\times}(x)$ as
\begin{equation}
h_{ij}(x)\, =\, h_{+}(x) e_{ij}^{+} + h_{\times}(x) e_{ij}^{\times}\, ,
\label{polarizedmetric}
\end{equation}
where 
\begin{align}
e^{+}=  \left(\begin{array}{cc}
1 & 0\\
0 & -1
\end{array}\right) \quad \quad \text{and}  \quad \quad 
 e^{\times}=  \left(\begin{array}{cc}
0 & 1\\
1 & 0
\end{array}\right).
\end{align}

Let us now consider the GWs propagating in a homogeneous, isotropic universe described by the FLRW metric 
\begin{equation}
\mathring{g}_{\mu\nu}dx^\mu dx^\nu = -N^2(x^0)\, d(x^0)^2 + a^2(x^0)d\mathbf{x}^2 ,
\label{metric0}
\end{equation}
where $x^0$ is an arbitrary time coordinate, $N(x^0)$ is the lapse function which depends on the choice of  $x^0$, and $d\mathbf{x}^2=\sum_{i}^3d(x^i)^2$ is a unit 3-sphere.
To study the linearized Einstein equations, and to be comparable with the Minkowski spacetime, it is more convenient to work with a conformally (perturbed) flat metric: 
\begin{equation}
g_{\mu\nu}\ =\ \mathring{g}_{\mu\nu} + h_{\mu\nu} = a^2\left(\eta_{\mu\nu}+\check{h}_{\mu\nu}\right).
\label{metric1}
\end{equation}
Here, the {\em conformal}  metric perturbation $\check{h}_{ij}$, for a wave traveling along the $x^3$ direction, is related to the physical metric perturbation (\ref{polarizedmetric}) by the scale factor as  
\begin{equation}
\check{h}_{ij}(x)\, :=\, a^{-2}h_{ij}(x) .
\label{polarizedmetric1a}
\end{equation}
The metric perturbation produces a perturbation to the  action  (\ref{Eq:EH-Action}). At second order in linear perturbations, for a traverse-traceless gauge, we get the perturbed action as \cite{Bardeen:1980kt} 
\begin{align}
\delta S_{\rm grav}^{(2)}\ =\ \frac{1}{4\kappa^2}\int d^4x  \sqrt{-\mathring{g}}\, 
\check{h}_{ij} \mathring{\Box} \check{h}^{ij} \, .
\label{Eq:EH-Action2}
\end{align}
This represents the action governing the GWs propagating on the unperturbed background  $\mathring{g}_{\mu\nu}$ in the $x^3$ direction.

For convenience,  let us introduce the new scalars $\check{h}_{\pm}(x)$  as 
\begin{equation}
\check{h}_{ij}(x) \, \coloneqq\, \sqrt{2}\kappa\left[\check{h}_{+}(x) e_{ij}^{+} + \check{h}_{\times}(x)e_{ij}^{\times}\right],
\label{polarizedmetric1}
\end{equation}
where
\begin{equation}
\check{h}_{+}(x) =  \frac{a^{-2}}{\sqrt{2}\kappa}\, h_{+}(x)  \quad \quad \text{and}\quad \quad  \check{h}_{\times}(x)  =  \frac{a^{-2} }{\sqrt{2}\kappa}\, h_{\times}(x)\, .
\end{equation}
By substitution Eqs.~(\ref{polarizedmetric}) and (\ref{polarizedmetric1}) into the perturbed action (\ref{Eq:EH-Action2}), 
the perturbed Lagrangian density at second order in linear perturbations
becomes
\begin{equation}
{\cal L}_{\check{h}}=\frac{1}{2}\sum_{\lambda =+,\times} \check{h}_{\lambda}\mathring{\Box} \check{h}_{\lambda}+{\cal O}(\check{h}_\lambda^{2}).
\label{eq:Lagrangian-Perturb}
\end{equation}
The effective action  of the independent polarization modes, provided by the Lagrangian density (\ref{eq:Lagrangian-Perturb}), is that of two massless scalar fields.
Thus, the equation of motion for the (scalar) perturbation $\check{h}_{\lambda}(x)$, with a fixed  $\lambda$, is given by the familiar Klein-Gordon equation
\begin{equation}
\mathring{\Box}\, \check{h}_{\lambda}(x) =0.
\label{Eq:Field}
\end{equation}
Henceforth, our aim will be to study the quantum theory of scalar perturbations $\check{h}_{\lambda}(x)$ satisfying the Klein-Gordon equation (\ref{Eq:Field}) propagating on the cosmological spacetime (\ref{metric0}).

The canonically conjugate pair for the  field $\check{h}_{\lambda}(x)$ consists of $(\check{h}_{\lambda}, \check{\pi}_{\lambda})$ on a $x^0=\mathrm{const.}$ slice.
As usual we would like to write the field $\check{h}_{\lambda}(x)$ in terms of
its Fourier modes. However, we are not \emph{a priori} assuming Lorentz
invariance and, in fact, we will be considering its possible violations.
Hence, we do not perform a four-dimensional Fourier transform on $\check{h}_{\lambda}(x)$;
rather we only consider such a transformation over spatial coordinates
for $\check{h}_{\lambda}(x)$ and its conjugate momentum $\check{\pi}_{\lambda}(x)$. 
The classical solutions of the equation of
motion (\ref{Eq:Field}) can be expanded in Fourier modes as 
\begin{subequations}
        \label{eq:H-lambda0}
\begin{align}
\check{h}_{\lambda}(x^0,\mathbf{x})\,  &=\, \frac{1}{\ell^{3/2}}\sum_{\mathbf{k}\in\mathscr{L}}\mathfrak{h}_{\lambda,\mathbf{k}}(x^0)e^{i\mathbf{k}\cdot\mathbf{x}},\label{eq:H-lambda}\\
\check{\pi}_{\lambda}(x^0,\mathbf{x})\, &=\,  \frac{1}{\ell^{3/2}}\sum_{\mathbf{k}\in\mathscr{L}}\Pi_{\lambda,\mathbf{k}}(x^0)e^{i\mathbf{k}\cdot\mathbf{x}},\label{eq: pi-lambda}
\end{align}
\end{subequations}
where the wave vector $\mathbf{k}\in(2\pi\mathbb{Z}/\ell)^{3}$ spans a
three-dimensional lattice\footnote{Because of the periodicity of the torus $\mathbb{T}^3$, equipped with coordinates $x^j\in(0, \ell)$,  the  allowed Fourier components are those with the wave vectors in the reciprocal space of $\mathbb{T}^3$. We therefore consider  an elementary cell $\mathcal{V}$ by fixing a fiducial (spatial) flat metric $^oq_{ij}$ and denote by $V_o=\ell^3$ the volume of  $\mathcal{V}$ in this geometry. (For simplicity, we  assume that this cell is cubical with respect to $^oq_{ij}$.) Then, all integrations in the Fourier expansion will be restricted to this volume.} $\mathscr{L}$ \citep{Ashtekar:2009mb}.  
The Fourier coefficients
are canonically conjugate satisfying the commutation relations $\{\mathfrak{h}_{\lambda,\mathbf{k}},\Pi_{\lambda,\mathbf{k}^{\prime}}\}=\delta_{\mathbf{k},-\mathbf{k}^{\prime}}$.
Moreover, the reality conditions on the field $h_{\lambda}(x^0,\mathbf{x})$  imply that
$\mathfrak{h}_{\lambda,\mathbf{k}}=(\mathfrak{h}_{\lambda,-\mathbf{k}})^{\ast}$
and $\Pi_{\lambda,\mathbf{k}}=(\Pi_{\lambda,-\mathbf{k}})^{\ast}$ 
are satisfied for each mode.

From the Lagrangian \eqref{eq:Lagrangian-Perturb}, we can write the
(time-dependent) Hamiltonian of the perturbation field propagating on the background $(M, \mathring{g}_{\mu\nu})$. In terms of the conjugate pairs $(\check{h}_{\lambda}, \check{\pi}_{\lambda})$,
by using Eqs.~\eqref{eq:H-lambda0}-\eqref{eq: pi-lambda}, the
Hamiltonian of the GW is obtained as 
\begin{align}
H(x^0)\, &= \, \sum_{\lambda=+,\times}\frac{N(x^0)}{2a^3(x^0)} \int_{\mathcal{V}} d^{3}x~\Big[(\check{\pi}_{\lambda})^{2}+a^4(x^0)(\partial_{i}\check{h}_{\lambda})^{2}\Big]\nonumber \\
&= \,  \frac{N(x^0)}{2a^3(x^0)}\sum_{\mathbf{k}}\sum_{\lambda=+,\times}\Big[\big(\Pi_{\lambda,\mathbf{k}}\big)^{\ast}\Pi_{\lambda,\mathbf{k}}+k^{2}a^4(x^0)\big(\mathfrak{h}_{\lambda,\mathbf{k}}\big)^{\ast}\mathfrak{h}_{\lambda,\mathbf{k}}\Big],\label{eq:Hamilton-app1}
\end{align}
where $k=|\mathbf{k}|$.

Following the above reality conditions for the perturbation field $h_{\sigma}(x^0,\mathbf{x})$,
it turns out  that not all modes $\mathfrak{h}_{\lambda,\mathbf{k}}(x^0)$
of the GWs are independent. In other words, when decomposing each
field mode $\mathfrak{h}_{\lambda,\mathbf{k}}(x^0)$ and its conjugate
momentum $\Pi_{\lambda,\mathbf{k}}(x^0)$ as 
\begin{align}
\mathfrak{h}_{\sigma,\mathbf{k}} &\coloneqq  \frac{1}{\sqrt{2}}\big(\mathfrak{h}_{\sigma,\mathbf{k}}^{(1)}+i\mathfrak{h}_{\sigma,\mathbf{k}}^{(2)}\big),\label{app-phi-pi-1a}\\
\Pi_{\sigma,\mathbf{k}} &\coloneqq  \frac{1}{\sqrt{2}}\big(\Pi_{\sigma,\mathbf{k}}^{(1)}+i\Pi_{\sigma,\mathbf{k}}^{(2)}\big),\label{app-phi-pi-1b}
\end{align}
the reality conditions imply that 
\begin{align}
\mathfrak{h}_{\sigma,-\mathbf{k}}^{(1)} &=  \mathfrak{h}_{\sigma,\mathbf{k}}^{(1)}, & \mathfrak{h}_{\sigma,-\mathbf{k}}^{(2)} &=  -\mathfrak{h}_{\sigma,\mathbf{k}}^{(2)}\\
\Pi_{\sigma,-\mathbf{k}}^{(1)} &=  \Pi_{\sigma,\mathbf{k}}^{(1)}, & \Pi_{\sigma,-\mathbf{k}}^{(2)} &=  -\Pi_{\sigma,\mathbf{k}}^{(2)}
\end{align}
For each $\mathbf{k}=(k_{1},k_{2},k_{3})$, the relation above enables
us to split the lattice $\mathscr{L}$ into positive and negative sectors
\citep{Ashtekar:2009mb} 
\begin{subequations}
\begin{align}
\mathscr{L}_{+} &=  \{\mathbf{k}:k_{3}>0\}\cup\{\mathbf{k}:~k_{3}=0,k_{2}>0\}\cup\{\mathbf{k}:k_{3}=k_{2}=0,k_{1}>0\}, \\
\mathscr{L}_{-} &= \ \{\mathbf{k}:k_{3}<0\}\cup\{\mathbf{k}:k_{3}=0,k_{2}<0\}\cup\{\mathbf{k}:k_{3}=k_{2}=0,k_{1}<0\}\nonumber \\
  &=\ \{\mathbf{k}:-\mathbf{k}\in\mathscr{L}_{+}\},
\end{align}
\end{subequations}
 respectively. This decomposition of $\mathscr{L}$ 
further enables us to decompose any summation over $\mathbf{k}\in\mathscr{L}$ into
its positive and negative parts. Then, we define the new variables ${\cal A}_{\lambda,\mathbf{k}}$
and ${\cal E}_{\mathbf{\lambda,k}}$, 
\begin{subequations}
\begin{align}
{\cal A}_{\lambda,\mathbf{k}} &\coloneqq \begin{cases}
\mathfrak{h}_{\lambda,\mathbf{k}}^{(1)}, & \textrm{for}\quad\mathbf{k}\in\mathscr{L}_{+}\\
\mathfrak{h}_{\lambda,-\mathbf{k}}^{(2)}, & \textrm{for}\quad\mathbf{k}\in\mathscr{L}_{-}
\end{cases}\label{def-q}\\
{\cal E}_{\mathbf{\lambda,k}} &\coloneqq  \begin{cases}
\Pi_{\lambda,\mathbf{k}}^{(1)}, & \textrm{for}\quad\mathbf{k}\in\mathscr{L}_{+}\\
\Pi_{\lambda,-\mathbf{k}}^{(2)}, & \textrm{for}\quad\mathbf{k}\in\mathscr{L}_{-}
\end{cases}\label{def-p}
\end{align}
\end{subequations}
 which are canonically conjugate
\begin{equation}
\left\{ {\cal A}_{\lambda,\mathbf{k}},{\cal E}_{\mathbf{\lambda^{\prime},k}^{\prime}}\right\} =\delta_{\mathbf{k}\mathbf{k}^{\prime}}\delta_{\lambda\lambda^{\prime}}.
\label{eq:PB-AE}
\end{equation}
Now, we can reexpress the Hamiltonian (\ref{eq:Hamilton-app1}) as 
\begin{equation}
H(x^0)\, =\, \frac{N}{2a^3}\sum_{\lambda=+,\times}\sum_{\mathbf{k}\in\mathscr{L}}\left[{\cal E}_{\mathbf{\lambda,k}}^{2}+k^{2}a^{4}{\cal A}_{\lambda,\mathbf{k}}^{2}\right]
\, =:\, \sum_{\lambda=+,\times}\sum_{\mathbf{k}\in\mathscr{L}} H_{\lambda, \mathbf{k}}(x^0).
\label{eq:Hamiltonian-FLRW-1}
\end{equation}
Equation~\eqref{eq:Hamiltonian-FLRW-1}  represents the Hamiltonian
of a set of decoupled harmonic oscillators defined by conjugate pairs
$({\cal A}_{\lambda,\mathbf{k}},{\cal E}_{\mathbf{\lambda,k}})$ associated
with any $\mathbf{k}$ mode for a fixed polarization $\lambda$, satisfying the relation \eqref{eq:PB-AE}.

At this point, we choose the harmonic time gauge where $N(x^0=\tau)=a^{3}(\tau)$
to get rid of the factor $a^{-3}$ in front of Eq. \eqref{eq:Hamiltonian-FLRW-1}.
Hence, the Hamiltonian of the perturbations (for the fixed mode $\mathbf{k}$ and polarization $\lambda$) over the FLRW background
in harmonic time becomes
\begin{equation}
H_{\lambda, \mathbf{k}}(\tau)\, =\, \frac{1}{2}\left[{\cal E}_{\mathbf{\lambda,k}}^{2}+k^{2}a^{4}{\cal A}_{\lambda,\mathbf{k}}^{2}\right].\label{eq:Hamiltonian-FLRW-2}
\end{equation}
This Hamiltonian, Eq. \eqref{eq:Hamiltonian-FLRW-2}, resembles an oscillator with time-dependent frequency, and therefore, analyzing its effective polymer quantum corrections is very complicated. The reason for this is that its polymer quantization will yield a time-dependent quantum pendulum-type system whose solutions are mathematically difficult to treat. In the next section we will show how we bypass this problem and obtain an effective polymer time-dependent Hamiltonian. 

\section{Polymer quantization and the effective Hamiltonian\label{sec:polymerization}}

As mentioned in the previous section, the Hamiltonian \eqref{eq:Hamiltonian-FLRW-1}
is a time-dependent one which makes finding its effective counterpart
complicated. In order to circumvent this issue, we will apply a procedure based on the extended phase space formalism (more details in Ref. \cite{garcia2017dirac}). The idea of the procedure is as follows. First, lift the system
to the extended phase space (EPS). In this way, time can now be considered as an additional degree of freedom at the price that instead of a true Hamiltonian we now have a first class constrained system, that is to say, a deparametrized time-dependent harmonic oscillator. Second, we apply a canonical transformation in the extended phase space in such a way that the time dependency of the Hamiltonian, when written in the new variables, is removed. As a result, once the constraint is fixed, we obtain a time-independent harmonic oscillator which can be polymerically quantized. At this point, the effective polymer terms arising in the semiclassical description are known \cite{austrich2017instanton, Tecotl:2015cya, Morales-Tecotl:2016dma}. We then consider these terms in what is now a polymer effective time-independent Hamiltonian and proceed to lift the system back to the extended phase space. Finally, we apply the inverse of the canonical transformation and solve the constraint. This yields the polymer effective Hamiltonian on the usual phase space, where now the Hamiltonian is not just effective but also time-dependent.

A schematic of our method can be seen in Fig. \ref{fig:schem}. The steps are written below or close to the arrows in parentheses, i.e., ``to EPS'' is
step (1), etc. In the following sections we detail this procedure.
Section \ref{subsec:TI-C-H} is devoted to steps (1) and (2), Sec.
\ref{subsec:PQ-TI-EH} discusses step (3), and in Sec. \ref{subsec:PEffTDH}
we will follow steps (4) and (5).

\tikzstyle{block} = [rectangle, draw, fill=blue!20, 
text width=7.5em, text centered, rounded corners, minimum height=3em]
\tikzstyle{line} = [draw, -latex']
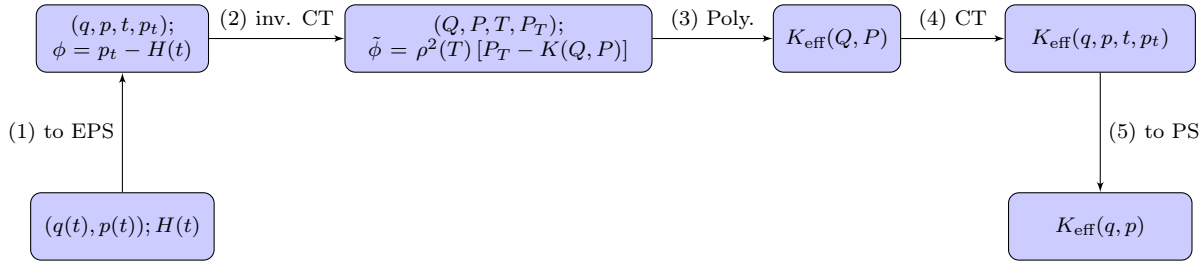
\begin{figure}[!t]
  \centering
  {\scriptsize
  \begin{tikzpicture}[node distance = 2cm, auto]
      
    \node [block] (consent) { $(q(t),p(t)); H(t)$};
    
    \node [block, above of = consent, node distance = 2.5cm,text width=7em] (screening) {$(q,p,t,p_t)$;\\ $\phi=p_t-H(t)$};
    
    \node [block, right of = screening, node distance = 5cm,text width=13em] (refer) {$(Q,P,T,P_T)$;\\ $\tilde{\phi}=\rho^{2}(T)\left[P_{T}-K(Q,P)\right]$};
    
    \node [block, right of = refer, node distance = 4.5cm,text width=5em] (refer2) { $K_{\textrm{eff}}(Q,P)$};
    
    \node [block, right of = refer2, node distance = 3.5cm,text width=7.8em] (refer3) { $K_{\textrm{eff}}(q,p,t,p_t)$};
    
   \node [block, below of = refer3, node distance = 2.5cm] (refer4) { $K_{\textrm{eff}}(q,p)$};
  
    \path [line] (consent) -- node {\textrm{(1) to EPS}}(screening);
    \path [line] (refer) -- node { \textrm{(3) Poly.}}(refer2); 
    \path [line] (refer2) -- node { \textrm{(4) CT}}(refer3);
    \path [line] (screening) -- node { \textrm{(2) inv. CT}}(refer);
    \path [line] (refer3) -- node {(5) to PS}(refer4);
     
  \end{tikzpicture}}
  \caption{Schematics of the derivation of a time-dependent effective Hamiltonian constraint. Here ``EPS'' means extended phase space, ``inv. CT'' denotes inverse canonical transformation, ``Poly.'' means the process of polymer quantization and getting an effective polymer Hamiltonian from there, ``CT'' denotes the canonical transformation, and ``PS'' means the nonextended phase space. The lower row corresponds to the usual phase space, while the upper row corresponds to the extended phase space. \label{fig:schem}}
\end{figure}

\subsection{Obtaining a time-independent classical Hamiltonian\label{subsec:TI-C-H}}

Let us consider a time-dependent harmonic oscillator 
\begin{equation}
S=\int\left\{ p\frac{dq}{dt}-H(t)\right\} dt,\label{TDHO-1}
\end{equation}
where the time-dependent Hamiltonian is of the form 
\begin{equation}
H(t)=\frac{1}{2m}p^{2}+\frac{1}{2}m\omega(t)^{2}q^{2}.
\end{equation}

\noindent We can now move to the extended phase space, step (1) in
Fig. \ref{fig:schem}, in which time $t$ is now one of the configuration
variables whose conjugate is denoted by $p_{t}$. Hence, the system
is now described by the coordinates $(q,t,p,p_{t})$. In accordance
with Dirac's formalism, the system is now described by the extended
action 
\begin{equation}
S=\int\left\{ p\frac{dq}{d\tau}+p_{t}\frac{dt}{d\tau}-\lambda\phi\right\} d\tau,\label{TDHOExt-1}
\end{equation}
where
\begin{equation}
\phi=p_{t}+H(t)\approx0,
\end{equation}
is a first class constraint ensuring the compatibility of the two
actions (\ref{TDHO-1}) and (\ref{TDHOExt-1}) in the usual and the extended phase space on the constrained surface $\phi=0$, and $\lambda$ is a Lagrange multiplier fixed to one once the constraint is solved. 

\noindent Next, in step 2 in Fig. \ref{fig:schem}, we perform
(the inverse of) a canonical transformation, 
\begin{align}
Q &=  \frac{1}{\rho(t)}q,\label{eq:can-tr-1}\\
T &=  \int\frac{1}{\rho^2(t)}\,dt,\label{eq:can-tr-2}\\
P &=  \rho(t)p-m\dot{\rho}(t)q,\label{eq:can-tr-3}\\
P_{T} &=  \rho^{2}(t)p_{t}+\rho(t)\dot{\rho}(t)\,q\,p  -\frac{m}{2}q^{2}\left[\dot{\rho}^{2}(t)+\frac{W^{2}}{\rho^{2}(t)}-\omega^{2}(t)\rho^{2}(t)\right],\label{eq:can-tr-4}
\end{align}
in order to transform the $H(t)$ appearing in the first class constraint
$\phi$ into a time-independent one. Here, $W$ is the time-independent
frequency of the time-independent system as we will see in Eq. \eqref{NewHo-1}
and $\rho$ is an auxiliary variable to be determined by the specific
properties of the system, more precisely by $\omega$ and $W$. 

Such a canonical transformation turns the action \eqref{TDHOExt-1} into
\begin{equation}
S=\int\left\{ P\frac{dQ}{d\tau}+P_{T}\frac{dT}{d\tau}-\lambda\tilde{\phi}\right\} d\tau,\label{TDHONew-1}
\end{equation}
where, the first class constraint now reads
\begin{equation}
\tilde{\phi}=\rho^{2}(T)\left[P_{T}+K\right]\approx0,\label{eq:phi-tild-class}
\end{equation}
and the corresponding Hamiltonian $K$ appearing in it is 
\begin{equation}
K=\frac{1}{2m}P^{2}+\frac{1}{2}mW^{2}Q^{2}.\label{NewHo-1}
\end{equation}

\noindent Moreover, the auxiliary equation used to fix $\rho(t)$
becomes 
\begin{equation} \label{eq:rho-eom}
\ddot{\rho}(t)+\omega^{2}(t)\rho(t)=\frac{W^{2}}{\rho^{3}(t)}.
\end{equation}
Now one can polymer quantize the time-independent Hamiltonian \eqref{NewHo-1}
as usual, find its effective counterpart, and then apply the canonical
transformations \eqref{eq:can-tr-1}--\eqref{eq:can-tr-4} to obtain
its associated extended action similar to Eq. \eqref{TDHOExt-1}, and
from there read off the time-dependent Hamiltonian in the usual (nonextended)
phase space. These are steps 3--5 in Fig. \ref{fig:schem}. These steps
will be detailed in the following subsections. Before continuing,
notice that in our paper the following correspondence holds
\begin{equation}
{\cal A}_{\sigma,\mathbf{k}}\to Q\quad \text{and} \quad {\cal E}_{\mathbf{\sigma,k}}\to P.\label{RelationQPAE}
\end{equation}

\subsection{Polymer quantization and effective time-independent Hamiltonian \label{subsec:PQ-TI-EH}}

\noindent Let us consider a time-independent Hamiltonian of the form
\eqref{NewHo-1} where the Poisson algebra of the canonical variables
is given by 
\begin{equation}
\{Q,P\}=1,\label{eq:QP-PB-sch}
\end{equation}
with other Poisson brackets being zero. Such a Poisson bracket allows
us to construct the Weyl algebra ${\cal W}$ whose generators $\widehat{W}(a,b)$
satisfy the Weyl algebra multiplication 
\begin{equation}
\widehat{W}\left(a_{1},b_{1}\right)\widehat{W}\left(a_{2},b_{2}\right)=e^{\frac{i}{2\hbar}\left(a_{1}b_{2}-b_{1}a_{2}\right)}\widehat{W}\left(a_{1}+a_{2},b_{1}+b_{2}\right),\label{eq:WA-stdrd}
\end{equation}
where $a_i$'s and $b_i$'s (with  $i=1, 2$) are parameters labeling the algebra generator $\widehat{W}$.
An example is the standard or Schr\"{o}dinger representation, where
the Weyl algebra ${\cal W}$ generators can be written as the formal
exponential 
\begin{equation}
\widehat{W}(a,b)=e^{\frac{i}{\hbar}(a\widehat{Q}-b\widehat{P})}. \label{WeylAlgGen}
\end{equation}

If the infinitesimal generators $\hat{Q},\hat{P}$ are both well defined on the Hilbert space, i.e. the conditions of the Stone-von Neumann theorems hold, then the Weyl algebra multiplication can be essentially reduced to $[\hat{Q},\hat{P}]=1$ of the Schr\"{o}dinger representation. However, we would like to perform a different quantization of our
classical system, known as the ``polymer representation''. As we will
see, in this type of quantization motivated by loop quantum gravity,
usually at least one of the infinitesimal generators $\hat{Q}$ or
$\hat{P}$ are not well defined on the Hilbert space due to the lack of
weak continuity of the operators (see below). This makes the polymer
representation unitarily inequivalent to the standard
Schr\"{o}dinger representation, and hence it is expected to yield
different physical results.

In polymer quantization one of the two fundamental operators, $\widehat{Q}$ or $\widehat{P}$, cannot be represented as an infinitesimal operator for the Weyl algebra generator as in Eq. (\ref{WeylAlgGen}). Moreover, the spectrum of the canonically conjugate variable is discrete. For example, if $\hat{Q}$ is not well defined, then the spectrum of its conjugate variable $\hat{P}$ becomes discrete. This is basically
because there is no $\hat{Q}$ to generate infinitesimal transformations
in $\hat{P}$. Naturally, the inverse of this statement is valid for the case
where $\hat{P}$ is not well defined. However, it is worth  noting that in LQG, the connection is holonomized/polymerized and the triad is discretized. Now, in our notation $Q$ corresponds to ${\cal A}_{\sigma,\mathbf{k}}$ which itself corresponds to the
metric perturbations; see Eq. \eqref{def-q}. Hence a representation where
$P$ or ${\cal E}_{\mathbf{\sigma,k}}$ is polymerized which results
in $Q$ or ${\cal A}_{\sigma,\mathbf{k}}$ becoming discrete is more
in line with LQG. In this work we will consider both cases  (i)  polymer $P$ and discrete $Q$, and (ii) polymer $Q$ and discrete $P$ in Secs. \ref{Case(i)} and \ref{Case(ii)} respectively.

\subsubsection{Case (i): Polymer $P$, discrete $Q$}  \label{Case(i)}

\noindent In this case, the polymer Hilbert space is of the form 
\begin{equation}
\mathscr{H}_{\textrm{poly}}^{(p)}=L^{2}\left(\overline{\mathbb{R}},dP_{Bohr}\right)\ni\Psi(P)=\sum_{\{Q_{j}\}}\Psi_{Q_{j}}e^{\frac{i}{\hbar}Q_{j}P},
\end{equation}
where $\overline{\mathbb{R}}$ is the Bohr compactification of the real line \cite{velhinho2007quantum} and $dP_{\mathrm{Bohr}}$ is the Bohr measure. The set of points $\{Q_{j}\}$, thought of as a graph, are discrete values corresponding to $Q$ and the inner product is 
\begin{equation}
\langle\Psi(P)|\Phi(P)\rangle=\lim_{L\rightarrow\infty}\frac{1}{2L}\int_{-L}^{L}\Psi(P)^{*}\Phi(P)dP.
\end{equation}

\noindent The representation of the Weyl algebra generators on $\mathscr{H}_{\textrm{poly}}^{(p)}$
is given by 
\begin{equation}
\widehat{W}(a,b)\Psi(P)=e^{\frac{i}{2\hbar}ab}e^{\frac{i}{\hbar}bP}\Psi(P+a).
\end{equation}
In this scheme, the operator $\widehat{W}(0,b)$ is not weakly continuous
\begin{equation}
\langle e^{\frac{i}{\hbar}Q_{j}P}|\widehat{W}(0,b)|e^{\frac{i}{\hbar}Q_{j}P}\rangle=\delta_{b,0},
\end{equation}
and consequently, it violates the Stone-von Neumann theorem requirements
for this representation to be unitarily equivalent to the standard
(Schr\"{o}dinger representation of) quantum mechanics. As a result,
we cannot obtain an infinitesimal generator for the operator $\widehat{W}(0,b)$ 
which, in the standard Schr\"{o}dinger representation corresponds
with $\widehat{P}$. For this reason, in polymer quantum mechanics,
we are forced to introduce a combination of Weyl generators that
mimics the term $\widehat{P}^{2}$ in the quantum Hamiltonian. In
order to introduce such a combination, the so-called \textit{polymer
scale} is needed. This scale, denoted by $\mu$ mimics the role of
the Planck length in LQG. While this is a free parameter
of the theory that should be fixed by experiment, it should be small enough to provide a good agreement with the experiments in standard quantum mechanics ($\mu/l_{0}\sim10^{-7}$, where $l_{0}$
is the proper length scale of the standard quantum harmonic oscillator).
Therefore, this polymer scale admits an upper bound. One way to put
a bound on the value of this scale is via the comparison of predicted
theoretical effects of polymer quantum mechanics on the propagation
of a GW and the experimental observations. This is
part of the motivation for the present work. 

Let us then consider a polymer scale $\mu$ with a fixed,
albeit unknown, value. Using $\mu$, the standard combination of Weyl
generators to provide the analog of $\widehat{P}^{2}$ is given by
\begin{equation}
\widehat{P}_{\textrm{poly}}^{2}=\frac{\hbar^{2}}{\mu^{2}}\left[2\widehat{1}-\widehat{W}(0,\mu)-\widehat{W}(0,-\mu)\right].
\end{equation}
As a result, the action of this operator is 
\begin{equation}
\widehat{P}_{\textrm{poly}}^{2}\Psi(P)=\left[\frac{2\hbar}{\mu}\sin\left(\frac{\mu P}{2\hbar}\right)\right]^{2}\Psi(P).
\end{equation}
It can be checked \citep{austrich2017instanton} that the in the semiclassical
limit, this operator yields the following expression for the quadratic
term ${P}^{2}$ in the Hamiltonian 
\begin{equation}
P_{\textrm{eff}}^{2}=\left[\frac{2\hbar}{\mu}\sin\left(\frac{\mu{P}}{2\hbar}\right)\right]^{2}.
\end{equation}
Using this result, the effective Hamiltonian for a polymer quantized
harmonic oscillator is of the form 
\begin{equation}
K_{\textrm{eff}}^{(p)}=\frac{1}{2m}\left[\frac{2\hbar}{\mu}\sin\left(\frac{\mu{P}}{2\hbar}\right)\right]^{2}+\frac{mW^{2}}{2}Q^{2}.\label{EffectiveHP}
\end{equation}

\subsubsection{Case (ii): Polymer $Q$, discrete $P$}  \label{Case(ii)}

In this case we can follow the same lines as in case (i). The
Hilbert space is now given by 
\begin{equation}
{\cal H}_{\textrm{poly}}^{(q)}=L^{2}\left(\overline{\mathbb{R}},dQ_{Bohr}\right)\ni\Psi(Q)=\sum_{\{P_{j}\}}\Psi_{P_{j}}e^{\frac{i}{\hbar}P_{j}Q},
\end{equation}
and the inner product is 
\begin{equation}
\langle\Psi(Q)|\Phi(Q)\rangle=\lim_{L\rightarrow\infty}\frac{1}{2L}\int_{-L}^{L}\Psi^{*}(Q)\Phi(Q)\,dQ.
\end{equation}

\noindent The representation for the Weyl generator in this Hilbert
space is 
\begin{equation}
\widehat{W}(a,b)\Psi(Q)=e^{-\frac{i}{2\hbar}ab}e^{-\frac{i}{\hbar}aQ}\Psi(Q+b).
\end{equation}
Note that the polymer scale in this case has units of $P$ and thus
we will use a different notation, $\nu$, for the polymer scale in
this case. In this representation the operator $\widehat{Q}$ is not
well defined and hence the term $\widehat{Q}^{2}$ in the Hamiltonian
is to be expressed using a combination of Weyl generators. The combination
is similar to the one considered for case (i), 
\begin{equation}
\widehat{Q}_{\textrm{poly}}^{2}=\frac{\hbar^{2}}{\nu^{2}}\left[2\widehat{1}-\widehat{W}(\nu,0)-\widehat{W}(-\nu,0)\right],
\end{equation}
and it can be checked that the action of this operator is 
\begin{equation}
\widehat{Q}_{\textrm{poly}}^{2}\Psi(Q)=\left[\frac{2\hbar}{\nu}\sin\left(\frac{\nu Q}{2\hbar}\right)\right]^{2}\Psi(Q)
\end{equation}

\noindent Similarly, the effective correction to the potential of
the harmonic oscillator is then given by 
\begin{equation}
Q_{\textrm{eff}}^{2}=\left[\frac{2\hbar}{\nu}\sin\left(\frac{\nu Q}{2\hbar}\right)\right]^{2},
\end{equation}
and the effective Hamiltonian in this case turns out to be 
\begin{equation}
K_{\textrm{eff}}^{(q)}=\frac{1}{2m}P^{2}+\frac{mW^{2}}{2}\left[\frac{2\hbar}{\nu}\sin\left(\frac{\nu Q}{2\hbar}\right)\right]^{2}.\label{EffectiveHQ}
\end{equation}

\subsection{Polymer time-dependent effective Hamiltonian\label{subsec:PEffTDH}}

After obtaining $\tilde{\phi}$, Eq. \eqref{eq:phi-tild-class} from
step (2), we can fix it using $\tilde{\phi}=0$ and $dT/d\tau=\dot{T}$
to obtain the Hamiltonian \eqref{NewHo-1}. In step (3), this time-independent
Hamiltonian is polymerized (as discussed in Sec. \ref{subsec:PQ-TI-EH}), from which an effective Hamiltonian is derived in the form
of either Eq. \eqref{EffectiveHP} or Eq. \eqref{EffectiveHQ}, depending on the
representation. This time-independent effective polymer Hamiltonian
is then replaced back into one of the following extended phase space
actions 
\begin{align}
S^{(P)}= & \int\left\{ P\frac{dQ}{d\tau}+P_{T}\frac{dT}{d\tau}-\tilde{\lambda}\tilde{\phi}^{(P)}\right\} d\tau,\\
S^{(Q)}= & \int\left\{ P\frac{dQ}{d\tau}+P_{T}\frac{dT}{d\tau}-\tilde{\lambda}\tilde{\phi}^{(Q)}\right\} d\tau.
\end{align}
based on the representation used. Next, in step (4), we perform the
canonical transformations \eqref{eq:can-tr-1}--\eqref{eq:can-tr-4}
on the above action, and particularly $\tilde{\phi}$, to obtain $\tilde{\phi}^{(p)}$
or $\tilde{\phi}^{(q)}$ as a function of extended phase space variables $q,t,p,p_{t}$.
It is worth noting that the canonical transformation introduces a
boundary term, which at the classical level does not affect the Lagrangian
equations of motion. 

Finally, in step (5), we solve the constraint  $\tilde{\phi}\approx0$
to obtain the time-dependent Hamiltonian in the usual phase space of $\left(q,p\right)$. Thus, one obtains the effective time-dependent polymer Hamiltonians
\begin{align}
H_{\textrm{eff}}^{(p)} &=  \frac{2\hbar^{2}}{m\mu^{2}\rho^{2}}\sin^{2}\left(\frac{\mu(\rho p-m\dot{\rho}q)}{2\hbar}\right)+\frac{\dot{\rho}q\,p}{\rho} +\frac{mq^{2}}{2}\left[\omega^{2}-\frac{\dot{\rho}^{2}}{\rho^{2}}\right],\label{eq:Hp-eff}\\
H_{\textrm{eff}}^{(q)} &=  \frac{p^{2}}{2m}+\frac{2m\hbar^{2}}{\nu^{2}}\left(\rho\ddot{\rho}+\omega^{2}\rho^{2}\right)\sin^{2}\left(\frac{\nu q}{2\hbar\rho}\right) -\frac{mq^{2}\ddot{\rho}}{2\rho}. \label{eq:Hq-eff}
\end{align}
The effective equations of motion corresponding to $H_{\textrm{eff}}^{(p)}$
are
\begin{align}
\frac{dq}{dt} &=  \left\{ q,H_{\textrm{eff}}^{(p)}\right\} =\frac{1}{m\rho}\frac{\hbar}{\mu}\sin\left(\frac{\mu\left(p\rho-mq\dot{\rho}\right)}{\hbar}\right)+\frac{\dot{\rho}(t)}{\rho(t)} q\, ,\\
\frac{dp}{dt} &=  \left\{ p,H_{\textrm{eff}}^{(p)}\right\} 
=  \frac{\dot{\rho}}{\rho^{2}}\frac{\hbar}{\mu}\sin\left(\frac{\mu\left(p\rho-mq\dot{\rho}\right)}{\hbar}\right)+\frac{mq\dot{\rho}^{2}}{\rho^{2}}-m\omega^{2}q- \frac{\dot{\rho}(t)}{\rho(t)} p\, ,
\end{align}
and the ones corresponding to $H_{\textrm{eff}}^{(q)}$ are
\begin{align}
\frac{dq}{dt} &=  \left\{ q,H_{\textrm{eff}}^{(q)}\right\} =\frac{p}{m},\\
\frac{dp}{dt} &=  \left\{ p,H_{\textrm{eff}}^{(q)}\right\} 
=  -m\hbar\left(\ddot{\rho}+\rho\omega^{2}\right)\frac{\sin\left(\frac{\nu q}{\hbar\rho}\right)}{\nu}+\frac{mq\ddot{\rho}}{\rho}.
\end{align}

\section{Effective equations of motion and phenomenology\label{sec:phenomenology}}

The correspondence between the generic analysis of the previous section
and our specific model is expressed as
\begin{align}
q &\to  \mathcal{A}_{\sigma,\mathbf{k}}, & p &\to  \mathcal{E}_{\sigma,\mathbf{k}},\label{eq:corres-1}\\
W^{2} &= \left|\mathbf{k}\right|^{2}, & \omega^{2} &= \left|\mathbf{k}\right|^{2}a^{4}, & m= & 1.\label{eq:corres-2}
\end{align}
Using these, we will study two effective descriptions of our model in what follows.

\subsection{Polymer $\mathcal{E}$, discrete $\mathcal{A}$}

By applying Eqs. \eqref{eq:corres-1}--\eqref{eq:corres-2} to Eq. \eqref{eq:Hp-eff}
we obtain the effective polymer Hamiltonian with polymer $\mathcal{E}_{\sigma,\mathbf{k}}$
as

\begin{align}
H_{\textrm{eff}}^{(\mathcal{E})}=\sum_{\lambda=+,\times}\sum_{\mathbf{k}\in\mathscr{L}} & \left\{ \frac{2}{\mu^{2}\rho^{2}}\sin^{2}\left(\frac{\mu\left(\rho\mathcal{E}_{\sigma,\mathbf{k}}-\dot{\rho}\mathcal{A}_{\sigma,\mathbf{k}}\right)}{2}\right)+\frac{\dot{\rho}\mathcal{A}_{\sigma,\mathbf{k}}\,\mathcal{E}_{\sigma,\mathbf{k}}}{\rho}+\frac{\mathcal{A}_{\sigma,\mathbf{k}}^{2}}{2}\left[\omega^{2}-\frac{\dot{\rho}^{2}}{\rho^{2}}\right]\right\} ,\label{eq:H-E-eff}
\end{align}
where we have set $\hbar=1$. The corresponding equations of motion
read 
\begin{align}
\frac{d{\cal A}_{\sigma,\mathbf{k}}}{dt}  &= \frac{1}{\rho}\frac{\sin\left(\mu\left(\rho\mathcal{E}_{\sigma,\mathbf{k}}-\dot{\rho}\mathcal{A}_{\sigma,\mathbf{k}}\right)\right)}{\mu} + \frac{\dot{\rho}}{\rho} \mathcal{A}_{\sigma,\mathbf{k}} ,\label{eq:EoM-eff-E-1}\\
\frac{d{\cal E}_{\sigma,\mathbf{k}}}{dt} &= \frac{\dot{\rho}}{\rho^{2}}\frac{\sin\left(\mu\left(\rho\mathcal{E}_{\sigma,\mathbf{k}}-\dot{\rho}\mathcal{A}_{\sigma,\mathbf{k}}\right)\right)}{\mu}+\left(\frac{\dot{\rho}}{\rho}\right)^{2}\mathcal{A}_{\sigma,\mathbf{k}}-\omega^{2}\mathcal{A}_{\sigma,\mathbf{k}} - \frac{\dot{\rho}}{\rho} \mathcal{E}_{\sigma,\mathbf{k}}.\label{eq:EoM-eff-E-2}
\end{align}
These equations are nonlinear in both ${\cal A}_{\sigma,\mathbf{k}}$
and $\mathcal{E}_{\sigma,\mathbf{k}}$, and their $\mu\to0$ limit
matches the classical equations of motion as expected. 

\subsection{Polymer $\mathcal{A}$, discrete $\mathcal{E}$}

In this case, and by applying Eqs. \eqref{eq:corres-1}--\eqref{eq:corres-2}
to Eq. \eqref{eq:Hq-eff} we obtain an effective polymer Hamiltonian with
polymer $\mathcal{A}_{\sigma,\mathbf{k}}$ as

\begin{align}
H_{\textrm{eff}}^{(\mathcal{A})}=\sum_{\lambda=+,\times}\sum_{\mathbf{k}\in\mathscr{L}} & \left\{ \frac{\mathcal{E}_{\sigma,\mathbf{k}}^{2}}{2}+\frac{2}{\nu^{2}}\left(\rho\ddot{\rho}+\omega^{2}\rho^{2}\right)\sin^{2}\left(\frac{\nu\mathcal{A}_{\sigma,\mathbf{k}}}{2\rho}\right)-\frac{\mathcal{A}_{\sigma,\mathbf{k}}^{2}\ddot{\rho}}{2\rho}\right\} .\label{eq:H-A-eff}
\end{align}
The equations of motion in this case are
\begin{align}
\frac{d{\cal A}_{\sigma,\mathbf{k}}}{dt} &=  \mathcal{E}_{\sigma,\mathbf{k}},\label{eq:EoM-eff-A-1}\\
\frac{d{\cal E}_{\sigma,\mathbf{k}}\left(t\right)}{dt} &=  -\frac{\ddot{\rho}+\rho\omega^{2}}{\nu}
\sin\left(\frac{\nu\mathcal{A}_{\sigma,\mathbf{k}}}{\rho}\right)
+\frac{\ddot{\rho}}{\rho}\mathcal{A}_{\sigma,\mathbf{k}},\label{eq:EoM-eff-A-2}
\end{align}
which are now nonlinear only in both ${\cal A}_{\sigma,\mathbf{k}}$,
while their $\nu\to0$ limit also matches the classical equations
of motion.

\subsection{Perturbative and nonperturbative numerical solutions}

We can solve Eqs.~\eqref{eq:EoM-eff-E-1}--\eqref{eq:EoM-eff-E-2} and \eqref{eq:EoM-eff-A-1}--\eqref{eq:EoM-eff-A-2} for specific field-space configurations, both perturbatively, and numerically and nonperturbatively in order to compute exact solutions that can be compared to perturbative calculations. We will begin by looking at solutions with a time-independent background, for which $\rho = 1$ and $\dot{\rho}=\ddot{\rho}=0$. After gaining some insight in this setting, we examine the solutions for $\rho$ required to study behavior in a time-dependent background. We can obtain solutions in this case by applying the transformation given in Eqs. \eqref{eq:can-tr-1}-\eqref{eq:can-tr-3} to the time-independent solution, or by directly solving the time-dependent equations of motion.

The essential parameters we would like to vary  include the mode amplitude at some initial time $t_I$, $\mathcal{A}_I \equiv \mathcal{A}(t=t_I)$, momentum $\mathcal{E}_I \equiv \mathcal{E}(t=t_I)$, frequency $\omega$, and $\mu$ (or $\nu$).
We can reduce this parameter space by considering the physical behavior of the system, and making note of several rescalings the equations of motion are invariant under.
We first note that the equations of motion are highly analogous to the case of a physical pendulum, and will similarly result in periodic behavior, albeit with a different period.
We can therefore seek solutions with an initial amplitude $\mathcal{A}_I = 0$ without loss of generality.
We then note that the equations of motion and auxiliary equation are invariant under a rescaling of the frequency,
\begin{align}
t & \rightarrow k t  &
\mu & \rightarrow k \mu  &
\bar{\nu} & \rightarrow \nu \nonumber\\
 & &
\mathcal{A} & \rightarrow \mathcal{A} &
 \mathcal{E} & \rightarrow \mathcal{E}/k \,, \label{eq:k-rescaling}
\end{align}
and so it suffices to obtain solutions for a single frequency.

The equations of motion are also invariant under a rescaling of the initial momentum $\mathcal{E}_I$,
\begin{align}
t & \rightarrow t  &
\mu & \rightarrow  \mathcal{E}_I \mu  &
\nu & \rightarrow \mathcal{E}_I \nu \nonumber\\
 & &
\mathcal{A} & \rightarrow \mathcal{A}/\mathcal{E}_I &
 \mathcal{E} & \rightarrow \mathcal{E}/\mathcal{E}_I\,, \label{eq:E-rescaling}
\end{align}
and so we can take the initial momentum to be $\mathcal{E}_I = 1$ in numerical solutions, although we will leave this factor in later analytic expressions. The parameters $\mu$ and $\nu$ then determine the ``smallness'' of oscillations. We note that the equations are invariant under a similar rescaling of $\mu$ and $\nu$, and so we could equivalently choose to vary $\mathcal{E}_I$; the important thing is to vary one of these quantities, which will determine how ``small'' the oscillations are. In either case, we have reduced the parameter space to a simple one in which we can vary only $\mu$ and $\nu$.

\begin{figure}[tb]
\includegraphics[width=1.0\textwidth]{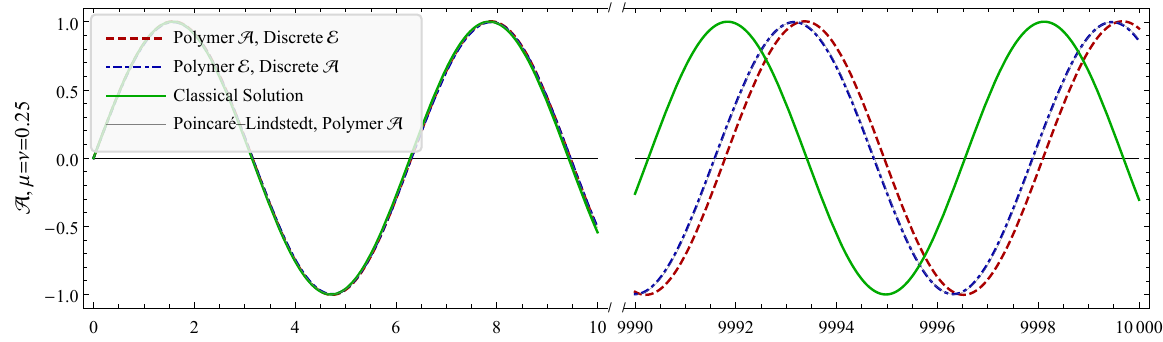}
\includegraphics[width=1.0\textwidth]{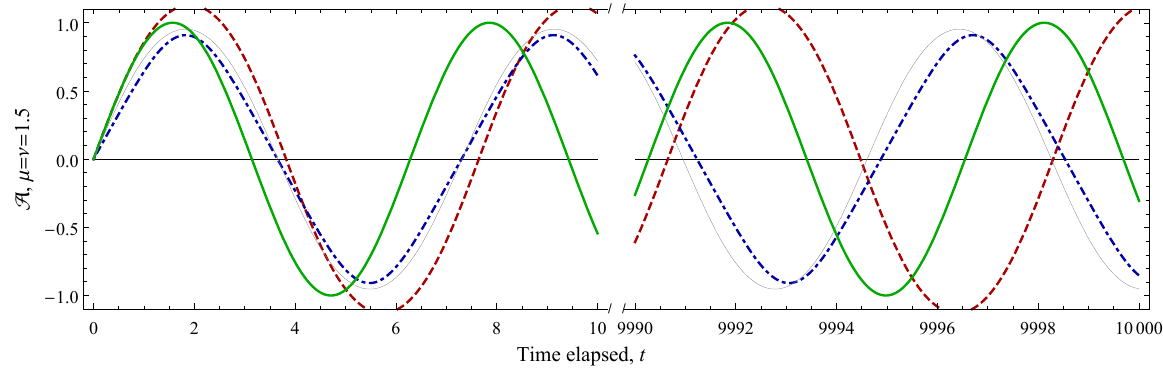}
\caption{ Time evolution of $\mathcal{A}$ with $\mathcal{A}_I=0$, $\mathcal{E}_I=1$, and $k=1$ for two different choices of $\mu = \nu$ in the case of a time-independent background spacetime, i.e., $\rho=$\, const. The solutions are shown at early times, and the axis is broken to show the behavior at a much later time. Solutions can be mapped to different choices of $k$ and $\mathcal{E}_I$ using the rescalings in Eqs.~\eqref{eq:k-rescaling} and~\eqref{eq:E-rescaling}, while changing $\mathcal{A}_I=0$ can be viewed as a phase shift. \label{fig:A-time-evol}}
\end{figure}

For the time-independent equations of motion (EoM), the solution for each wave vector is in fact identical to that of an ordinary physical pendulum for polymer $\mathcal{A}$. While solutions to this equation and the corresponding one for polymer $\mathcal{E}$ are periodic, due to the nonlinear structure both higher harmonics may be excited and a frequency shift develops. Both of these corrections are of order $\mathcal{O}(\nu^2)$ (or $\mu^2$). 
The frequency shift is not readily found using a standard perturbative approach, where the solution appears to contain a secular, growing term. However, this term can be eliminated by also expanding in a small perturbation of the frequency; this is the essence of the Poincar\'e-Lindstedt method, which we employ here to obtain an approximate analytic solution. %

For polymer $\mathcal{A}$, again fixing the phase so that $\mathcal{A}_I = 0$, the solution we obtain is given by 
\begin{equation}
\mathcal{A}(t) \simeq \mathcal{E}_I \sin\left[(1 - (\mathcal{E}_I \nu)^{2}/16) k t \right] - \frac{\mathcal{E}_I^3 \nu^{2}}{48} \sin^{3}\left[(1 - (\mathcal{E}_I \nu)^{2}/16) k t \right]\,,\label{eq:polyA-Abar-pert}
\end{equation}
while for polymer $\mathcal{E}$ the solution is 
\begin{align}
\mathcal{A}(t) \simeq & \,\, \mathcal{E}_I \sin\left[(1-(\mathcal{E}_I k \mu)^{2}/16) k t \right] \nonumber \\
 & -\frac{\mathcal{E}_I^3 k^2 \mu^2 }{16} \sin^{2}\left[(1-(\mathcal{E}_I k \mu)^{2}/16) k t \right] \cos\left[(1-(\mathcal{E}_I k \mu)^{2}/16) k t \right] \,.\label{eq:polyE-Abar-pert}
\end{align}
These solutions can be seen to contain a frequency shift of order $\nu^2$ or $\mu^2$, and a cubic correction term. The frequency shifts in both cases are nearly identical; this is because while the role of conjugate variables has been interchanged in the solutions, the form has remained unchanged. The second, cubic term can also be rewritten, and thought of, as an introduction of higher harmonics using angle identities. In observations, the frequency shift may be more important to account for than the excited harmonics. This is because the frequency shift can manifest as a phase shift that has considerable time to develop as the wave traverses cosmological distances. In Fig.~\ref{fig:A-time-evol} we demonstrate this, comparing the perturbative solution to the exact and classical ones for the time-independent case.

We can also analyze the above perturbative solutions and obtain some insight into the speed of propagation of the waves. For that, we note that the dominant contributions to Eqs.~\eqref{eq:polyA-Abar-pert} and \eqref{eq:polyE-Abar-pert}
can be written as
\begin{align}
\mathcal{A}(t) & \simeq  \mathcal{E}_{I}\sin\left[\left(1-\left(\frac{\mathcal{E}_{I}\nu}{4}\right)^{2}\right)kt\right],\\
\mathcal{A}(t) & \simeq \mathcal{E}_{I}\sin\left[\left(1-\left(\frac{\mathcal{E}_{I}k\mu}{4}\right)^{2}\right)kt\right].
\end{align}
Comparing with the classical solution where we identify $ka^{2}=\omega_{c}$, with $\omega_{c}$ being the classical angular speed,
we notice that up to first order the polymer angular speeds are
\begin{align}
\omega_{\nu}^{(\mathcal{A})}\,  & \simeq\, \omega_{c}\left[1-\left(\frac{\mathcal{E}_{I}\nu}{4}\right)^{2}\right],\label{eq:omega-nu-approx}\\
\omega_{\mu}^{(\mathcal{E})}\, & \simeq\, \omega_{c}\left[1-k^{2}\left(\frac{\mathcal{E}_{I}\mu}{4}\right)^{2}\right].\label{eq:omega-mu-approx}
\end{align}
Although these are perturbative and approximate and even though we
have neglected higher harmonics in Eqs. \eqref{eq:polyA-Abar-pert} and
\eqref{eq:polyE-Abar-pert}, the above two equations reveal a curious
phenomenon. Noting that $\omega_{c}=ka^{2}$ and with the group velocity
being 
\begin{equation}
v=\frac{d\omega_{\textrm{poly}}}{d\left(ka^{2}\right)}
\end{equation}
with $\omega_{\textrm{poly}}$ being either $\omega_{\nu}^{(\mathcal{A})}$
or $\omega_{\nu}^{(\mathcal{E})}$, we obtain
\begin{align}
v_{\nu}^{(\mathcal{A})}\,  &\simeq\, 1-\left(\frac{\mathcal{E}_{I}\nu}{4}\right)^{2},\label{eq:grp-v-A}\\
v_{\mu}^{(\mathcal{E})}\, & \simeq\, 1-k^{2}\left(\frac{\mathcal{E}_{I}\mu}{4}\right)^{2}.\label{eq:grp-v-E}
\end{align}
where $v_{\nu}^{(\mathcal{A})}$ and $v_{\mu}^{(\mathcal{E})}$ are
velocities of the effective waves in the case of polymer $\mathcal{A}$
and polymer $\mathcal{E}$, respectively. One can see from Eq. \eqref{eq:grp-v-A}
that in the polymer $\mathcal{A}$ case, the group velocity of the waves
is slower than the speed of light by a factor of $\left(\frac{\mathcal{E}_{I}\nu}{4}\right)^{2}$
that does not depend on the frequency of the waves, but is dependent
on the initial momentum $\mathcal{E}_{I}$ of the waves and the polymer
parameter, in this case $\nu$. Hence, all of the waves in this case move
slower than the speed of light and this  effect is amplified
if the wave has a larger initial momentum $\mathcal{E}_{I}$. For
the polymer $\mathcal{E}$ case in which we are more interested, we can
see from Eq. \eqref{eq:grp-v-E} that such a lower-than-the-speed-of-light
propagation also happens for the waves, and it also depends on the
initial momentum $\mathcal{E}_{I}$ of the waves and the polymer parameter
$\mu$ due to the factor $k^{2}\left(\frac{\mathcal{E}_{I}\mu}{4}\right)^{2}$. However, in this case there is an important difference: the
deviation from the speed of light also depends on the modes $k$.
Hence, waves with larger $k$ (i.e., larger energies) have a lower
speed compared to the ones with smaller $k$ and are more affected
by the quantum structure of spacetime. Also, notice that this case
leads to the violation of Lorentz symmetry as can be seen by squaring
both sides of Eq. \eqref{eq:omega-mu-approx}. Of course, due to the sheer
smallness of the expected values of $\mu$ and $\nu$, and the appearance of their
squares in the above expressions, these effects are very small, but
a highly energetic phenomenon with a large $\mathcal{E}_{I}$ may
help to amplify it to an extent that future observatories can detect it. We should emphasize that the presence of the violation of the Lorentz symmetry in this case, as seen from the above results, is a consequence of the polymer quantization and, in particular, this model, and is not a direct consequence of LQG. 

For the case of a time-dependent background, we can obtain a solution in one of two ways: directly integrating the EOMs, or using the canonical transformation in Eqs.~\eqref{eq:can-tr-1}--\eqref{eq:can-tr-4}. In either case, we will need to obtain a solution for $\rho$ by solving Eq.~\eqref{eq:rho-eom}. 
In general, this choice determines whether the mode amplitude will be purely decaying or will contain oscillatory behavior. Here we will seek purely growing solutions for $\rho$, choosing initial conditions such that oscillatory behavior is minimized; in our case, simply choosing $\rho = 1$ and $\dot{\rho} = 0$ is sufficient. Choosing a different initial amplitude for $\rho$ is in any case equivalent to rescaling of the scale factor $a$, polymer scale, momentum, and time coordinate.

\begin{figure}[tb]
    \centering
    \includegraphics[width=0.6\textwidth]{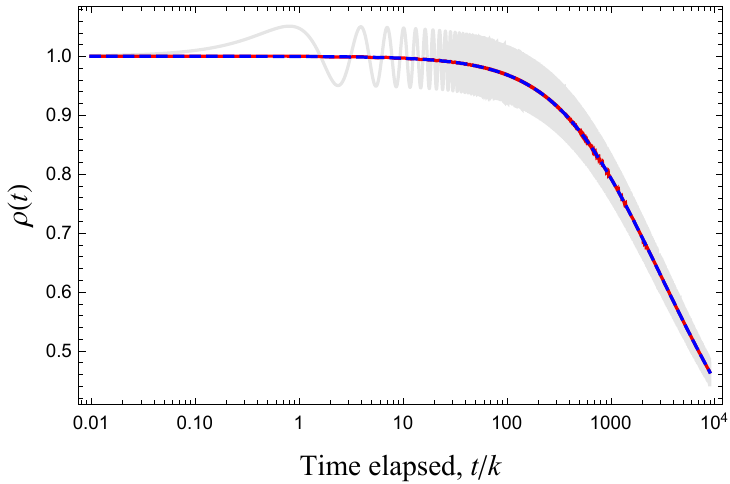}
    \caption{Evolution of the auxiliary variable $\rho(t)$. The full numerical nonoscillatory solution is shown in solid red, an  approximate power-law solution is shown in dashed blue, and a solution with initial conditions that result in oscillatory behavior is shown in light grey.}
    \label{fig:rho-t}
\end{figure}

For the case of a time-dependent background, the solutions can be obtained by transforming the ones with the time-independent background,
\begin{equation}
\mathcal{A}(t) \simeq \mathcal{E}_I \rho \sin\left[(1 - (\mathcal{E}_I \nu)^{2}/16) k T(t) \right] -  \frac{\mathcal{E}_I^3 \nu^{2}}{48} \rho \sin^{3}\left[(1 - (\mathcal{E}_I \nu)^{2}/16) k T(t) \right]\,,\label{eq:polyA-Abar-pert2}
\end{equation}
\begin{align}
\mathcal{A}(t) \simeq & \,\, \mathcal{E}_I \rho \sin\left[(1-(\mathcal{E}_I k \mu)^{2}/16) k T(t) \right] \nonumber \\
 & -\frac{\mathcal{E}_I^3 k^2 \mu^2 }{16} \rho \sin^{2}\left[(1-(\mathcal{E}_I k \mu)^{2}/16) k T(t) \right] \cos\left[(1-(\mathcal{E}_I k \mu)^{2}/16) k T(t) \right] \,,\label{eq:polyE-Abar-pert2}
\end{align}
where
\begin{equation}
    T(t) = \int_{t_I}^{t} dt' \frac{1}{\rho(t')^2}
\end{equation}
For GWs emitted at a time much greater than the characteristic wave time scale, ie., $t_I \gg k^{-1}$, and for nonoscillatory solutions, the second-derivative term is small, and solutions to the auxiliary equations are well approximated by a simple power law, $\rho = 1/a$.
In Fig.~\ref{fig:rho-t} we show the behavior of $\rho$ for several sets of initial conditions, and for a universe with a cosmological constant with $w=-1$, $a \propto t^{1/3}$, and $t_I = 10^{3}$ (in units of $k^{-1}$). In subsequent plots we will use initial conditions that do not result in oscillatory behavior.

\begin{figure}[tb]
\includegraphics[width=1.0\textwidth]{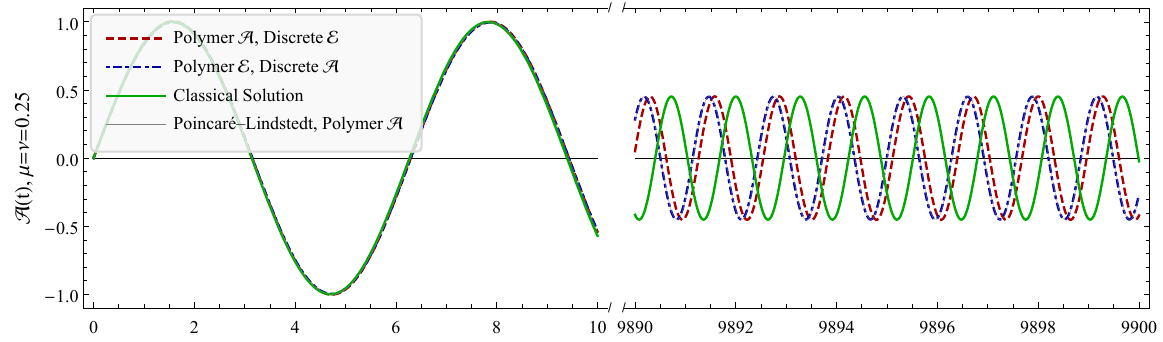}
\includegraphics[width=1.0\textwidth]{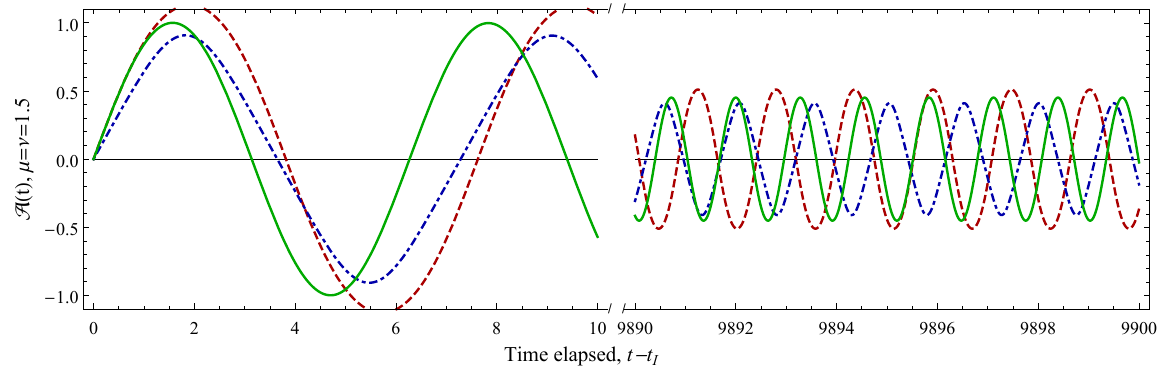}
\caption{Time evolution of $\mathcal{A}$ (as in Fig.~\ref{fig:A-time-evol})  for two different choices of $\mu = \nu$, for the case of a time-dependent background, i.e., $\rho(t)$ as described in the text. The axis is broken to show the behavior at a later time. \label{fig:A-time-evol-flrw}}
\end{figure}

From the canonical transformation~\eqref{eq:can-tr-1}--\eqref{eq:can-tr-3} (or, rather, its inverse), we see that the time-dependent waveform amplitude will pick up an overall factor of $\rho$ relative to the time-independent one, the time coordinate will be altered, and the momentum will be similarly rescaled but will also pick up an additional factor proportional to the wave amplitude. Due to the monotonically decreasing nature of $\rho$ and the smallness of its derivative, this additional factor will be a strongly subdominant contribution.
In Fig.~\ref{fig:A-time-evol-flrw} we show the final solution for the field $\mathcal{A}(t)$ for this time-dependent background.
Somewhat counterintuitively, the frequency is seen to increase at later times; more commonly the frequency is considered to decrease (redshift) with cosmological expansion. 
This is due to the choice of harmonic slicing we have made, with $N = a^3$ instead of the more commonly used $N=1$ (synchronous) or $N = a$ (comoving) time coordinate.

\section{Discussion and Conclusion\label{sec:conclusion}}
In this work we have studied a certain effective form of GWs, considered as quantized perturbations propagating over a classical FLRW spacetime, in order to derive observational signatures to be compared with the results of experiments conducted by GW observatories. We have considered the Hamiltonian of classical gravitational perturbations, a time-dependent Hamiltonian, and have applied the techniques of polymer quantization to it. Polymer quantization is a nonperturbative method of quantization, inspired by LQG, in which some of the operators are regularized and written in a certain exponential form. Since such a quantization is unitarily inequivalent to the standard Schr\"{o}dinger representation, one expects to obtain physically distinct results compared to it. We explored two polymer representations:   one in which the configuration variables are regularized (or polymerized) and their momenta are discretized, and one in which the momenta are polymerized and the configuration variables are discretized. We consider both representations for the polymer quantization of the aforementioned time-dependent Hamiltonian: one in which the perturbations are polymerized and their momenta are discrete, and in which the momenta of the perturbations are polymerized and hence the perturbations themselves are discretized. Of course the latter case is more interesting to us. 

It is worth mentioning that this polymer quantization was applied to each of the Fourier modes of the GW. A feature of this quantization is that the one-particle Hilbert space is modified and the Lorentz symmetry is no longer present \cite{garcia2016polymer}. This modification is ``encoded'' on each of the polymer scales ($\mu$ or $\nu$), which are usually considered to be very small (of the order of the Planck scale). However, our intuition in the present case is that the propagation of the GWs may capture some insights about these modifications despite the small values of the polymer scales.

Since the classical Hamiltonian we obtained is time dependent, to overcome the challenge of polymer quantizing such a time-dependent system we applied a method that is used to deal with the same issue in time-dependent harmonic oscillators \cite{garcia2017dirac}. We first wrote such a Hamiltonian in a time-independent format in the extended phase space by applying a certain canonical transformation, polymer quantized it, recovered its effective description, and applied the inverse of such a  canonical transformation to make it time dependent again. We finally transformed it back into the standard phase space obtaining in this way a time-dependent polymer quantized effective  Hamiltonian. We then derived and numerically studied the corresponding effective fully nonperturbative equations of motion. We also derived a perturbative analytical expression for the solutions and analyzed them to obtain further insight into the behavior of such waves. 
As a result, we found the following. 
\begin{enumerate}
    \item[i)] The form of the waves is modified. More precisely, there is a phase shift with respect to the classical case. Furthermore, small-amplitude harmonics are excited.
    \item[ii)] The speed of the waves turns out to be smaller than the speed of light. In a perturbative analysis, we find the following for the time-independent background (the same qualitative behavior is seen numerically for the time-dependent case): 
    \begin{enumerate} 
    \item In the case where the gravitational perturbations are polymerized and their conjugate momenta are discretized, the wave speeds are $v_{\nu}^{(\mathcal{A})} \simeq 1-\left(\frac{\mathcal{E}_{I}\nu}{4}\right)^{2}$. Hence, the factor $\left(\frac{\mathcal{E}_{I}\nu}{4}\right)^{2}$ by which the speed of waves differ from the speed of light depends on the polymer scale $\nu$ and the initial wave momentum $\mathcal{E}_{I}$, and this is the same for all of the waves regardless of their wave vectors or frequencies. Of course, this factor is very small due to the expected small value of the polymer parameter, in this case, $\nu$.
    \item In the case where the momenta of the perturbations are polymerized and the gravitational perturbations themselves are discretized (which is the more interesting case for us) the wave speeds are $v_{\mu}^{(\mathcal{E})} \simeq 1-k^{2}\left(\frac{\mathcal{E}_{I}\mu}{4}\right)^{2}$. Hence, in this case the factor $k^{2}\left(\frac{\mathcal{E}_{I}\mu}{4}\right)^{2}$ by which the wave speed is smaller than the speed of light not only depends on the polymer scale $\mu$ and the initial momentum of the perturbations $\mathcal{E}_{I}$, but now it also depends on the wave vector $k$ or, equivalently, the frequency of the waves. Thus, the higher-energy waves show a greater deviation from the classical behavior compared to the low-energy waves.
    \end{enumerate}
    \item[iii)] The modifications to the classical behavior due to quantum effects become increasingly visible as  the waves travel: the corrections result in an effective phase shift, which can become of order unity when $\mathcal{E}_I \mu^2 k^3 D_s$ or $\mathcal{E}_I \nu^2 k D_s$ are of order unity for a distance $D_s$ traveled.
\end{enumerate}

The power spectrum of primordial GWs originating from the Planckian era in the early Universe have been extensively explored in quantum gravity theories. In particular, in the context of LQC, various scenarios---such as the dressed metric  (see, e.g., Refs. \cite{Grain:2010yv,Agullo:2012sh,Agullo:2012fc, Agullo:2013ai}), deformed algebra (see, e.g., Refs. \cite{Grain:2010yv, Linsefors:2012et, Barrau:2014maa, Martineau:2017tdx}), and hybrid quantization approaches (see, e.g., Refs. \cite{FernandezMendez:2012vi, Gomar:2015oea, deBlas:2016puz, Gomar:2017yww})  approaches were employed to study the  power spectrum of the cosmological  perturbations. Therein,  deviations from standard general relativity in the sub-Planckian regimes have been investigated, which  led to observable signatures in the power spectrum of the cosmic microwave background. (For a comparison between these two approaches see,  e.g., Ref. \cite{Bolliet:2015bka}.)
Consequently, within our present setting and for a specific cosmological (or astrophysical) background,
when the mode function solutions $\mathcal{A}_{\sigma, \mathbf{k}}$ to the EoM  (\ref{eq:EoM-eff-E-1})--(\ref{eq:EoM-eff-E-2}) and (\ref{eq:EoM-eff-A-1})--(\ref{eq:EoM-eff-A-2}) are known, it is possible to  calculate  the primordial   (or nonprimordial) power spectra as ${\cal P}_{\mathcal{A}}(k) \sim (k^{3}/2\pi^{2})\,  |\mathcal{A}_{\sigma,\mathbf{k}}|^2$. Thereby, a  Polymer quantum-induced departure from the standard (quantum) theory of cosmological perturbations on a classical spacetime is obtained.  
We will address these subjects in detail in a companion paper, which is in preparation  \cite{Garcia-Chung:Progress}.
Furthermore, we plan to obtain a more robust constraint on $\mu$ and $\nu$ in future works, where
we will apply these results to initial data known from real GWs, and compare the numerical results of applying our method to waves with such initial values with the observed results of GW observatories, particularly those of LIGO. Furthermore, we will proceed to apply our method to the case where both the background spacetime and the perturbations are effective.

\begin{acknowledgments}
Y. T. and S. R. conducted this work within  the  Action CA18108--\emph{Quantum gravity phenomenology in the multi-messenger approach}--supported by the European Cooperation in Science and Technology (COST). This research was supported in part by Perimeter Institute for
Theoretical Physics. Research at Perimeter Institute is supported by
the Government of Canada through the Department of Innovation, Science
and Economic Development Canada and by the Province of Ontario through
the Ministry of Research, Innovation and Science. P. V. M. and Y. T.  acknowledge the FCT grants UIDBMAT/00212/2020 and UIDPMAT/00212/2020 at CMA-UBI.
\end{acknowledgments}

\appendix

\section{Friedmann equations in harmonic slicing}

In a majority of the cosmological literature, the slicing condition used either coincides with a synchronous time and corresponding lapse $N=1$, or conformal time with $N=a$. In this work we choose a harmonic slicing with $N=a^3$, which results in a modified behavior for the evolution of the scale factor. The coupled Einstein-fluid equations for a homogeneous, isotropic universe in a 3+1 language are given by
\begin{eqnarray}
\partial_t \ln \gamma & = & -2NK \\
\partial_t K & = & \frac{N}{3}K^2 + 4\pi N(\rho_m +3P) \\
\partial_t (a^3 \rho_m) & = & 0\,.
\end{eqnarray}
for spatial metric determinant $\gamma = a^6$, trace of the extrinsic curvature $K$, and Arnowitt-Deser-Misner (ADM) density and pressure $\rho_m$ and $P$. Assuming an equation of state $P=w\rho_m$ to close the system, and choosing the lapse $N=a^3$, this system has solutions of the form
\begin{equation}
    \label{eq:scalefactor}
    a(t) = \left( t/t_I \right)^{ \frac{2}{3} \frac{1}{1-w} }\,.
\end{equation}
For equations of state $w=-1,0,1/3$ (cosmological constant, dust, radiation), the scale factor shows power-law growth. We will eventually be interested in solving Eq.~\eqref{eq:rho-eom} for a given choice of $a$; in general, e.g., in a universe with multiple components, we will need to solve for $\rho$ numerically.

\bibliography{main}

\end{document}